# Detecting nonlinearity in multivariate time series

Milan Paluš

*Institute of Computer Science, Academy of Sciences of the Czech Republic*

*Pod vodárenskou věží 2, 182 07 Prague 8, Czech Republic; and*

*Santa Fe Institute, 1399 Hyde Park Road, Santa Fe, NM 87501, USA*

E-mail: `mp@uivt.cas.cz, mp@santafe.edu`

June 20, 1995

### Abstract

We propose an extension to time series with several simultaneously measured variables of the nonlinearity test, which combines the redundancy – linear redundancy approach with the surrogate data technique. For several variables various types of the redundancies can be defined, in order to test specific dependence structures between/among (groups of) variables. The null hypothesis of a multivariate linear stochastic process is tested using the multivariate surrogate data. The linear redundancies are used in order to avoid spurious results due to imperfect surrogates. The method is demonstrated using two types of numerically generated multivariate series (linear and nonlinear) and experimental multivariate data from meteorology and physiology.

## 1 Introduction

A number of methods for detecting nonlinearity and determinism in experimental time series have been proposed recently [1, 2, 3, 4, 5, 6, 7][1]. Input data, used by these methods, are usually univariate time series. In many real-world problems, however, multivariate time series are available; either several different variables are simultaneously[2] recorded from a system under study, like atmospheric pressure, temperature and humidity in meteorology; heart rate, blood pressure, respiration and blood oxygen saturation in physiology; mutual exchange rates of several currencies in economy; or, in spatially extended systems, multichannel recordings are obtained from probes at different spatial locations, like in studies of turbulent liquids, satellite data, electrocardiograms (ECG) or electroencephalograms (EEG). A few authors have proposed methods for studying nonlinear relations among variables [10, 11, 12], and multichannel EEG or ECG recordings were used as inputs into dimensional algorithms in some cases [13, 14, 15, 16, 17], however, majority of analyses of multivariate time series are usually based on linear methods like cross-correlations and orthogonal decomposition (principal component/factor analysis, Karhunen-Loève expansion), even in nonlinear dynamics context (see [16, 18] and references within).

Recently, Prichard & Theiler [19] have proposed an extension to multivariate time series of the phase-randomized Fourier-transform algorithm for generating surrogate data. In this letter we present a multivariate version of the method for testing nonlinearity, which combines an information-theoretic (redundancy) approach with the surrogate data technique. The univariate implementation of the method is described in detail in [6]. The use of the information-theoretic functionals, called redundancies, has at least three important advantages in comparison to other nonlinear statistics. First, various types of the redundancies can be constructed in order to test very specific types of dependence between/among (groups of) variables, including conditional relations. Second, the redundancies can be naturally evaluated as functions of time lags, so that dependence structures under study are not evaluated statically, but with respect to dynamics of a system under investigation. Third, for any type of the redundancy its linear form exists, which is sensitive to linear dependence only. These "linear redundancies" are used for testing quality of the surrogate data in order to avoid spurious detection of nonlinearity caused by imperfect surrogates.

---

[1] The references cited here are related to the area of nonlinear dynamics and chaos. A number of tests for (non)linearity have been developed in statistics (see, e.g., the review in [8]). On the other hand, the presented method and related tests are also generally applicable in time series analysis.

[2] In some cases synchronous measurement is impossible, however, the series can be synchronized in data preprocessing, see, e.g., [9].

## 2 Multivariate surrogate data

The basic idea in the surrogate-data based nonlinearity test is to compute a *nonlinear* statistic from data under study and from an ensemble of realizations of a linear stochastic process, which mimics "linear properties" of the studied data. If the computed statistic for the original data is significantly different from the values obtained for the surrogate set, one can infer that the data were not generated by a linear process; otherwise the null hypothesis, that a linear model fully explains the data, is accepted and the data can be further analyzed and characterized by using well-developed linear methods. For the purpose of such test the surrogate data must preserve the spectrum[3] and consequently, the autocorrelation function of the series under study. In the multivariate case also cross-correlations between all pairs of variables must be preserved. An isospectral linear stochastic process to a series can be constructed by computing the Fourier transform (FT) of the series, keeping unchanged the magnitudes of the Fourier coefficients, but randomizing their phases and computing the inverse FT into the time domain. Different realizations of the process are obtained using different sets of the random phases. In the multivariate case, the cross-correlations can be preserved by preserving phase differences between individual variables, i.e., the phases are randomized by adding random numbers, so that for a particular frequency bin the same random number is added to related phases of all the variables. More details can be found in [19] and references therein.

## 3 Entropy, information and redundancy

Let $X$ be a discrete random variable with a set of values $\Xi$ and probability mass function $p(x) = \Pr\{X = x\}$, $x \in \Xi$. We denote the probability mass function by $p(x)$, rather than $p_X(x)$, for convenience.

The *entropy* $H(X)$ of a discrete random variable $X$ is defined by

$$H(X) = -\sum_{x \in \Xi} p(x) \log p(x). \tag{1}$$

For a pair of discrete random variables $X$ and $Y$ with a joint distribution $p(x, y)$ the *joint entropy* $H(X, Y)$ is defined as

$$H(X, Y) = -\sum_{x \in \Xi} \sum_{y \in \Upsilon} p(x, y) \log p(x, y). \tag{2}$$

The *conditional entropy* $H(Y|X)$ of $Y$ given $X$ is defined as

$$H(Y|X) = \sum_{x \in \Xi} p(x) H(Y|X = x) = -\sum_{x \in \Xi} \sum_{y \in \Upsilon} p(x, y) \log p(y|x). \tag{3}$$

The average amount of common information, contained in the variables $X$ and $Y$, is quantified by the *mutual information* $I(X; Y)$, defined as

$$I(X; Y) = H(X) + H(Y) - H(X, Y). \tag{4}$$

The joint entropy of $n$ variables $X_1, \ldots, X_n$ with the joint distribution $p(x_1, \ldots, x_n)$ is defined as

$$H(X_1, \ldots, X_n) = -\sum_{x_1 \in \Xi_1} \ldots \sum_{x_n \in \Xi_n} p(x_1, \ldots, x_n) \log p(x_1, \ldots, x_n). \tag{5}$$

---

[3] Also, preservation of histogram is usually required. A histogram transformation used for this purpose is described in [6] and references within, and will not be discussed here, as far as its application is not different from the univariate case, it just should be applied to all variables individually.

*Redundancy* $R(X_1; \ldots ; X_n)$ quantifies the average amount of common information contained in the $n$ variables $X_1, \ldots, X_n$ and can be defined as straightforward generalization of (4):

$$R(X_1; \ldots ; X_n) = H(X_1) + \ldots + H(X_n) - H(X_1, \ldots, X_n). \tag{6}$$

Besides (6), the *marginal redundancy* $\varrho(X_1, \ldots, X_{n-1}; X_n)$, quantifying the average amount of information about the variable $X_n$ contained in the variables $X_1, \ldots, X_{n-1}$, can be defined as

$$\varrho(X_1, \ldots, X_{n-1}; X_n) = H(X_1, \ldots, X_{n-1}) + H(X_n) - H(X_1, \ldots, X_n). \tag{7}$$

The relation

$$\varrho(X_1, \ldots, X_{n-1}; X_n) = R(X_1; \ldots ; X_n) - R(X_1; \ldots ; X_{n-1}) \tag{8}$$

can be derived by simple manipulation.

In addition to the redundancy (6) and the marginal redundancy (7) we can define various types of redundancies quantifying the average amounts of information between/among variables or groups of variables. For instance, considering variables $X_1, \ldots, X_n, Y_1, \ldots, Y_m, Z_1, \ldots, Z_k$, the redundancy among the three groups of $X$'s, $Y$'s and $Z$'s is

$$R(X_1, \ldots, X_n; Y_1, \ldots, Y_m; Z_1, \ldots, Z_k) = H(X_1, \ldots, X_n) + H(Y_1, \ldots, Y_m)$$
$$+ H(Z_1, \ldots, Z_k) - H(X_1, \ldots, X_n, Y_1, \ldots, Y_m, Z_1, \ldots, Z_k). \tag{9}$$

Or, the redundancy between the $X$'s and $Y$'s (considered together) and the $Z$'s (a generalization of the marginal redundancy for groups of variables) is

$$R(X_1, \ldots, X_n, Y_1, \ldots, Y_m; Z_1, \ldots, Z_k) = H(X_1, \ldots, X_n, Y_1, \ldots, Y_m)$$
$$+ H(Z_1, \ldots, Z_k) - H(X_1, \ldots, X_n, Y_1, \ldots, Y_m, Z_1, \ldots, Z_k). \tag{10}$$

Conditional redundancies can be considered if relations between/among (groups of) variables are studied, in which influence of other (group of) variable(s) should be eliminated. For instance, the conditional redundancy between the groups of the $X$'s and the $Y$'s given the $Z$'s is

$$R(X_1, \ldots, X_n; Y_1, \ldots, Y_m | Z_1, \ldots, Z_k) = H(X_1, \ldots, X_n | Z_1, \ldots, Z_k)$$
$$+ H(Y_1, \ldots, Y_m | Z_1, \ldots, Z_k) - H(X_1, \ldots, X_n, Y_1, \ldots, Y_m | Z_1, \ldots, Z_k). \tag{11}$$

Or, the conditional redundancy among $X_1, \ldots, X_n$ given $Y_1, \ldots, Y_m$ is

$$R(X_1; \ldots ; X_n | Y_1, \ldots, Y_m) = H(X_1 | Y_1, \ldots, Y_m) + H(X_2 | Y_1, \ldots, Y_m) + \ldots$$
$$+ H(X_n | Y_1, \ldots, Y_m) - H(X_1, \ldots, X_n | Y_1, \ldots, Y_m). \tag{12}$$

Now, let the $n$ variables $X_1, \ldots, X_n$ have zero means, unit variances and correlation matrix $\mathbf{C}$. Then, we define the *linear redundancy* $L(X_1; \ldots ; X_n)$ of $X_1, X_2, \ldots, X_n$ as

$$L(X_1; \ldots ; X_n) = -\frac{1}{2} \sum_{i=1}^{n} \log(\sigma_i), \tag{13}$$

where $\sigma_i$ are the eigenvalues of the $n \times n$ correlation matrix $\mathbf{C}$.

If $X_1, \ldots, X_n$ have an $n$-dimensional Gaussian distribution, then $L(X_1; \ldots ; X_n)$ and $R(X_1; \ldots ; X_n)$ are theoretically equivalent [20].

Based on (8) we define the *linear marginal redundancy* $\lambda(X_1, \ldots, X_{n-1}; X_n)$, quantifying linear dependence of $X_n$ on $X_1, \ldots, X_{n-1}$, as

$$\lambda(X_1, \ldots, X_{n-1}; X_n) = L(X_1; \ldots ; X_n) - L(X_1; \ldots ; X_{n-1}). \tag{14}$$

Similarly, for any type of the general (nonlinear) redundancy its linear equivalent exists — it can be either obtained as a combination of the redundancies of the type (13) based on related relation to the redundancies of the type (6)[4], or derived directly using relevant expressions for multidimensional Gaussian distributions.

The general redundancies $R$ detect all dependences in data under study, while the linear redundancies $L$ are sensitive only to linear structures. (For detailed discussion see [6]. For more information about information-theoretic functionals see [21] and references in [4, 5, 6].)

## 4  The method

An experimentalist usually deals with a multivariate time series $\{x_1(t), \ldots, x_n(t)\}$, $t = 1, \ldots, N$, which is considered as a realization of a multivariate, stationary and ergodic stochastic process $\{X_1(t), \ldots, X_n(t)\}$. Then, due to ergodicity, all the information-theoretic functionals can be estimated using time averages instead of ensemble averages; in particular, correlation matrices in (13) are obtained as the time averages over the series, and probability distributions, used in computation of the redundancies $R$, are estimated as time-averaged histograms. When the discrete variables $X_1, \ldots, X_n$ are obtained from continuous variables on a continuous probability measure space, then the redundancies $R$ depend on a partition $\xi$ chosen to discretize the space. Various strategies have been proposed to define an optimal partition for estimating redundancies of continuous variables (see [4, 5, 6] and references therein). Here we use the "marginal equiquantization" method described in detail in [4, 5, 6], which is applied to each variable separately.

Having chosen an appropriate type of the redundancy, say $R(X_1; \ldots; X_n)$, it is possible and useful to investigate dependences among lagged series, i.e., to evaluate redundancies of the type $R(x_1(t); x_2(t + \tau_1); \ldots; x_n(t + \tau_{n-1}))$. Due to stationarity this redundancy does not depend on the time $t$ and is a function of the lags $\tau_1, \ldots, \tau_{n-1}$.

Like in [6] we define the test statistic as the difference between the redundancy obtained for the original data and the mean redundancy of a set of surrogates, in the number of standard deviations (SD's) of the latter. Thus both the redundancies and redundancy-based statistics are functions of the lags $\tau_1, \ldots, \tau_{n-1}$, and their graphs are $(n-1)$-dimensional hyperplanes. Such objects are hard to study and therefore it would be practically useful to define a few one-dimensional cuts of the hyperplanes. The cuts along the axes (i.e., setting all $\tau$'s but one to be equal to zero) are the simplest choice, more complicated cuts could be probably defined considering a problem under study[5]. Evaluating the redundancies and related statistics for broad ranges of the lags can bring a problem of simultaneous statistical inference (see [6, 7] and references within for details). This approach, however, can be more reliable than single-valued tests, as it was demonstrated in univariate case in [6]. As far as all the related considerations from [6] directly apply to multivariate problems, we refer readers to Ref. [6] and will not repeat them here, similarly as the discussion of the function of the linear redundancy, which is used to check the quality of the surrogate data: In some cases the surrogates can have auto-/cross-correlations different from the original data. This difference is detected by the redundancies $R$ (or other nonlinear statistics) and can be erroneously interpreted as detection of nonlinearity in linear data. The linear redundancy-based statistic evaluates just the differences in the "linear properties", i.e., in autocorrelations in the univariate tests and in crosscorrelations in the multivariate tests. A significant result in linear redundancy-based statistic indicates a problem in the surrogates and necessity of further investigation before a conclusion about (non)linearity of data under study is made [6].

---

[4] I.e., any redundancy can be expressed as a combination of the redundancies of the type (6). Such an expression always contains the redundancy of all variables under study, as far as an n-dimensional redundancy cannot be generally obtained from a sum of lower-dimensional redundancies.

[5] Testing nonlinearity in univariate series [6] also redundancies of several variables were evaluated. Those variables, however, were obtained using a one lagged variable, and $\tau_i = i\tau$, with one variable $\tau$, was a very natural choice. Studying multivariate series, however, this is not the case.

# 5 Examples

Multivariate time series obtained from a bivariate linear autoregressive (AR) model and the three-dimensional nonlinear chaotic Lorenz system are used as examples of numerically generated data of known origin in order to demonstrate the proposed method. Two real-data examples, using multivariate experimental time series from meteorology and physiology, are also presented.

## 5.1 Linear autoregression

16,384 samples of the bivariate series $\{x(t), y(t)\}$ were generated by the linear AR model:

$$x(t) = 0.9x(t-1) + \sigma_1(t),$$

$$y(t) = 0.9x(t-1) + 0.9y(t-1) + \sigma_2(t),$$

where $\sigma_1(t)$ and $\sigma_2(t)$ are Gaussian deviates with zero means and unit variances. The results – the linear redundancy $L(x(t); y(t+\tau))$, the redundancy $R(x(t); y(t+\tau))$, the linear (linear redundancy $L$-based) statistic and the nonlinear (redundancy $R$-based) statistic as functions of the time lag $\tau$ are presented in Fig. 1. The redundancies for the data and for the surrogates (Figs. 1a,b) coincide. Both the linear and nonlinear statistics (Fig. 1c and 1d, respectively) are confined between the values -1.6 and 1.6 SD's, i.e., the data are not significantly different from the surrogates. The linear stochastic hypothesis is accepted in agreement with the origin of the series.

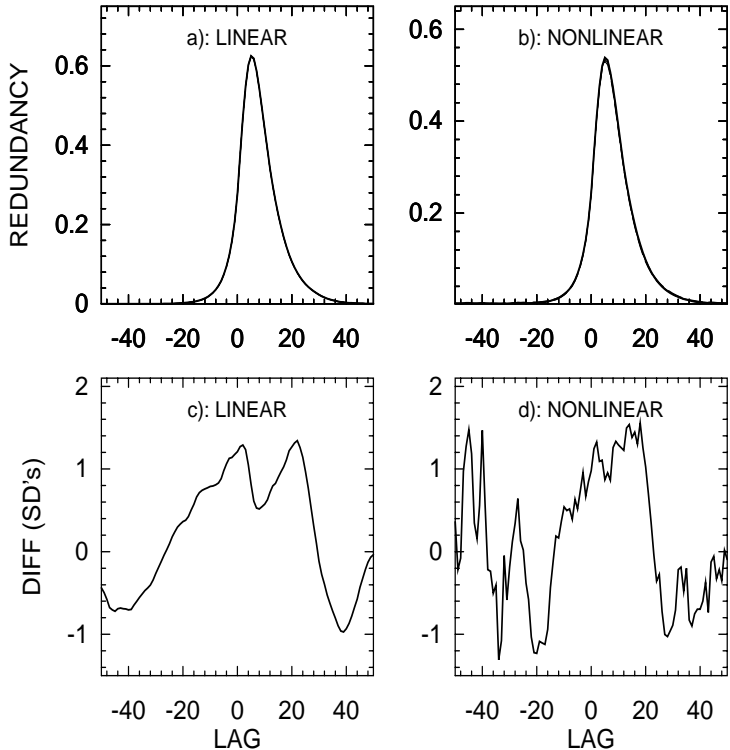

Figure 1: a): Linear redundancy $L(x(t); y(t+\tau))$, b): nonlinear (general) redundancy $R(x(t); y(t+\tau))$, for a bivariate linear autoregressive process and (coinciding curves) for related isospectral surrogates (mean of a set of 30 realization of the surrogates); c): linear (L-based), and d): nonlinear (R-based) statistics; as functions of the time lag $\tau$.

## 5.2 Lorenz system

16,384 samples of the three-variable series $\{x(t), y(t), z(t)\}$ were obtained from the chaotic Lorenz system [22]:
$$(dx/dt, dy/dt, dz/dt) = (10(y-x), 28x - y - xz, xy - 8z/3), \qquad (15)$$
integrated by the Bulirsch-Stoer method [23] with initial values (15.34, 13.68, 37.91), integration step 0.04 and accuracy 0.0001. Studying dependences of all the three variables, the redundancies $R(x(t); y(t+\tau_1); z(t+\tau_2))$, $L(x(t); y(t+\tau_1); z(t+\tau_2))$ and the related statistics should be evaluated. The results for the cut $\tau_2 = 0$ are presented in Fig. 2:

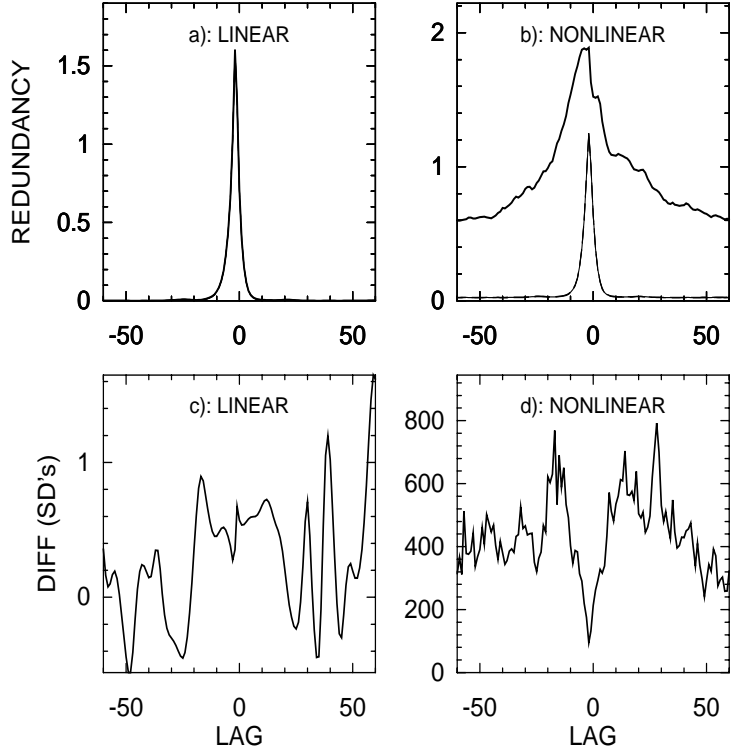

Figure 2: a): Linear redundancy $L(x(t); y(t+\tau_1); z(t))$, b): nonlinear (general) redundancy $R(x(t); y(t+\tau_1); z(t))$, for the chaotic three-dimensional Lorenz system (thick line) and for related isospectral surrogates (thin line, coinciding in a, but different in b); c): linear, and d): nonlinear statistics; as functions of the time lag $\tau_1$.

$L(x(t); y(t+\tau_1); z(t))$ for the data and the surrogates (Fig. 2a) coincide, the values of the linear statistic are between -0.5 and 1.6 SD's. Thus the *linear* dependence structures in the data are not different from those in the surrogates, i.e., the surrogates are technically good and should not be a source of spurious results in the test. The redundancy $R(x(t); y(t+\tau_1); z(t))$ (Fig. 2b) for the data (thick, upper curve) is clearly different[6] from the mean redundancy $R(x(t); y(t+\tau_1); z(t))$ for the surrogates (thin, lower curve), and the nonlinear statistic (Fig. 2d) indicates highly significant differences (hundreds of SD's). Thus the linear stochastic null hypothesis is rejected and, considering also results from linear statistic, nonlinearity is reliably detected.

---
[6] Note that $L(x(t); y(t+\tau_1); z(t))$ for both the data and surrogates and $R(x(t); y(t+\tau_1); z(t))$ for the surrogates have the same shape, while the shape of $R(x(t); y(t+\tau_1); z(t))$ of the data is different. See the discussion on "qualitative testing" in [6], which is also relevant for multivariate tests.

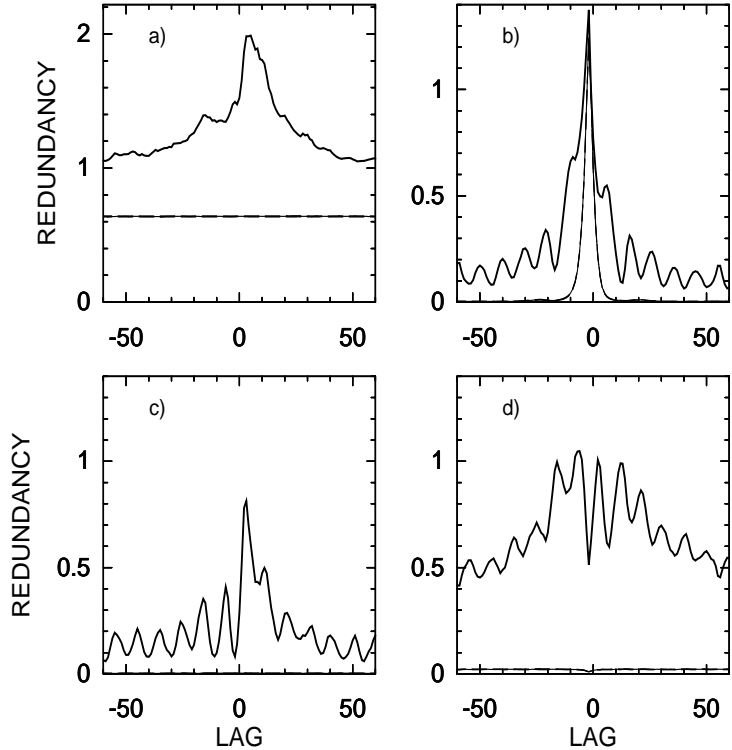

Figure 3: a): Nonlinear (general) redundancy $R(x(t); y(t); z(t+\tau_2))$, b): $R(x(t); y(t+\tau_1))$, c): $R(x(t); z(t+\tau_2))$, and d): marginal redundancy $\varrho(x(t), y(t+\tau_1); z(t))$; as functions of the time lag $\tau_1$ or $\tau_2$, for the chaotic three-dimensional Lorenz system (thick lines) and for related isospectral surrogates (thin lines, coinciding with the abscissa in c and d).

Other dependence structures in the three Lorenz variables are studied in Fig. 3[7]. The three-variable redundancy $R(x(t); y(t); z(t+\tau_2))$ (i.e., the cut $\tau_1 = 0$, perpendicular to that in Fig. 2b) is displayed in Fig. 3a. $R(x(t); y(t); z(t+\tau_2))$ of the data (thick, upper curve) is again clearly different from $R(x(t); y(t); z(t+\tau_2))$ of the surrogates, which is constant. This can be understood from the other results in Fig. 3:

$R(x(t); y(t+\tau_1))$ (Fig. 3b) shows that linear dependence between $x$ and $y$ ($R(x(t); y(t+\tau_1))$ of the surrogates, thin curve) is limited to small range of $\tau$'s around zero, while nonlinear dependence ($R(x(t); y(t+\tau_1))$ of the data, thick curve) is spread to larger lags. This result also shows the usefulness of evaluation of the redundancies and the test statistics for a large range of lags: In lags equal or close to zero the linear dependence is similar to the nonlinear dependence (in the sense of small difference between $R(x(t); y(t+\tau_1))$ of the data and of the surrogates) and a test limited to these lags could neglect nonlinearity. Evaluating the test for larger lags the differences are clear and highly significant (Fig. 3b).

Figure 3c illustrates $R(x(t); z(t+\tau_2))$ of the data (thick curve). $R(x(t); z(t+\tau_2))$ of the surrogates is close to zero and coincide with the abscissa, indicating that $x$ and $z$ are linearly independent[8]. Relation between $y$ and $z$ is very similar.

---

[7] The complete test result consists of the four graphs of $L$ and $R$ and the related statistics like in Figs. 1 and 2. Here we present only the graphs of the redundancies $R$. The remaining three graphs, for all the four cases in Fig. 3, can be characterized as follows: $L$ of the data does not differ from $L$ of the surrogates and they both look like $R$ of the surrogates; linear statistic is nonsignificant, while differences in the nonlinear statistic are highly significant (hundreds of SD's).

[8] In fact, there is a very weak linear dependence between $x$ and $z$ (or $y$ and $z$), which can be confirmed using the linear statistic on so called scrambled surrogates. This relation, however, is practically negligible when compared to the strong nonlinear dependence between $x$ and $z$, or to the linear dependence between $x$ and $y$. Therefore we use the above formulation about independence. The same statement holds for the results in Fig. 3d.

Marginal redundancy $\varrho(x(t), y(t+\tau_1); z(t))$, indicating the dependence of $z$ on both $x$ and $y$, is displayed in Fig. 3d: The thick line is for $\varrho$ of the data, $\varrho$ of the surrogates coincides with the zero line. This finding, i.e., linear independence of $z$ from $x$ and $y$ explains the result in Fig. 3a: The only contribution to the linear dependence (preserved in the surrogates) in the three-variable $R(x(t); y(t+\tau_1); z(t+\tau_2))$ is from the variables $x$ and $y$ and depends on $\tau_1$, but is independent of $\tau_2$. Therefore $R(x(t); y(t+\tau_1); z(t+\tau_2))$ of the surrogates depends only on $\tau_1$ and is constant in slices $\tau_1 = $ const., as in Fig. 3a.

## 5.3 Meteorological data

Daily recordings of atmospheric surface temperature and pressure $\{T(t), P(t)\}$ ($t = 1 - 65{,}536$ days; i.e., 180 years) were analyzed individually in [7]. The temperature series was found consistent with the null hypothesis of a linear stochastic process, while a significant nonlinear component was detected in the pressure series. Here we study the relationship between the two series $\{P(t)\}$ and $\{T(t)\}$ using the redundancies $L(P(t); T(t+\tau))$, $R(P(t); T(t+\tau))$ and the related statistics.

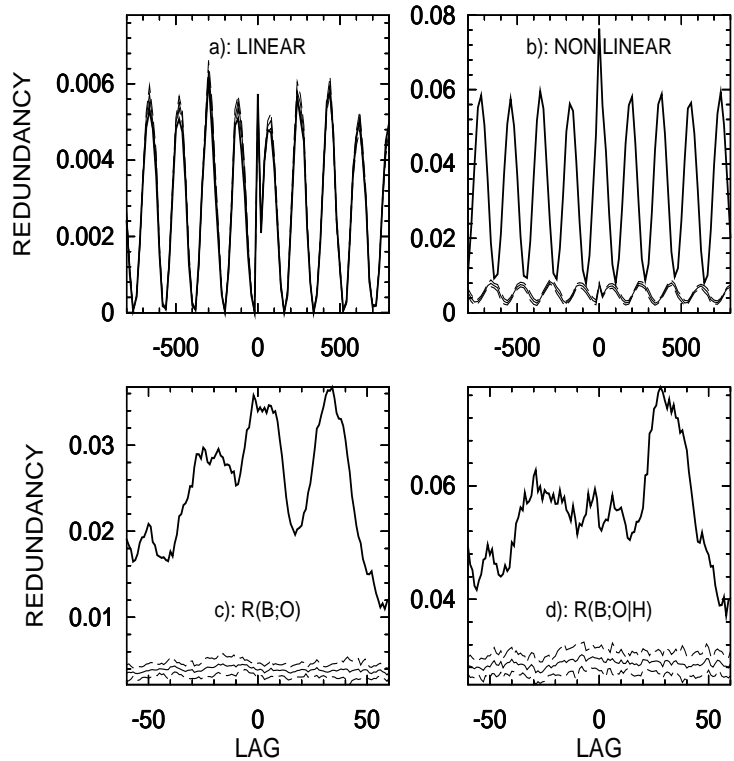

Figure 4: a): Linear redundancy $L(P(t); T(t+\tau))$, b): nonlinear (general) redundancy $R(P(t); T(t+\tau))$, for the bivariate meteorological series (mean daily atmospheric pressure and temperature); c): redundancy $R(B(t); O(t+\tau))$ for the physiological variables ($B$ – respiration, $O$ – blood oxygen saturation), and d): the conditional redundancy $R(B(t); O(t+\tau)|H(t))$ between $B$ and $O$ given the heart rate series $H$; as functions of the lag $\tau$. In all the cases the redundancies of the original data are drawn by thick lines, thin lines are used for mean values of the surrogate sets, dashed thin lines present values of mean $\pm$ SD for the surrogates. The lags $\tau$ are in days in a) and b), and in 0.5 sec. samples in c) and d).

The linear redundancy $L(P(t); T(t+\tau))$ for the data and for the surrogates coincide (Fig. 4a), also, there is no significant difference in the linear statistic (here we do not display the statistics in order to save space), i.e., the surrogates mimic well the linear dependences of the series. The redundancy $R(P(t); T(t+\tau))$ (Fig.

4b), however, shows striking differences between the data and the surrogates. In both $R(P(t); T(t+\tau))$ from the data and $R(P(t); T(t+\tau))$ from the surrogates a yearly periodicity is apparent, there is, however, a large difference in values (leading to differences of 150 SD's in the nonlinear statistics), and the oscillations in linear and nonlinear dependence (as reflected in the redundancies) are not in phase, but are shifted by approximately 75 (or, -107) days. Considering the above classification of the individual series, a rather naive interpretation of this result can be that the pressure series is as a two-component function of the (linear) temperature series. The first component of the function is nonlinear and the second one is linear and delayed 107 days with respect to the former. More realistic model should consider a common periodic driving for both the series. Detailed discussion and interpretations, however, are beyond the scope of this paper, and will be published elsewhere, together with results of further testing.

## 5.4 Physiological data

A multivariate physiological series, recorded from a patient in a sleep laboratory, was a part of the data used in the Santa Fe Institute Time Series Contest [18]. Here we have chosen the following three variables: instantaneous heart rate $\{H(t)\}$, a respiratory variable $\{B(t)\}$ quantifying breathing (a function of air flow passing nostrils), and the blood oxygen saturation $\{O(t)\}$. The segment processed here, chosen visually to be as stationary as possible, starts at sample 5500 of B1.dat and consists of 8192 samples (see Fig. 5 in [9], p. 120). The sampling time is 0.5 sec.

The redundancy $R(B(t); O(t+\tau))$, quantifying the dependence between the respiration and the blood oxygen, is displayed in Fig. 4c. The distinction between the data and the surrogates is clear, i.e., nonlinearity in this relation is clearly detected. Let us focus our attention on $R(B(t); O(t+\tau))$ itself. It detects peaks in both positive and negative lags, i.e., $\{B(t)\}$ influences $\{O(t)\}$, but also $\{O(t)\}$ influences $\{B(t)\}$. This can be an inherent relation, but also an effect of other variables in the system. In particular, there is a nonlinear relation between $\{O(t)\}$ and $\{H(t)\}$ and both linear and nonlinear dependences between $\{B(t)\}$ and $\{H(t)\}$. The conditional redundancy $R(B(t); O(t+\tau)|H(t))$ of $\{B(t)\}$ and $\{O(t)\}$ given $\{H(t)\}$ (Fig. 4d) should evaluate the relation between the respiration $\{B(t)\}$ and the blood oxygen saturation $\{O(t)\}$, eliminating the effect of the heart rate $\{H(t)\}$. Although $R(B(t); O(t+\tau)|H(t))$ is positive for both negative and positive lags, the principal peak is located in lags $26 - 30$; i.e., due to a nonlinear mechanism, a change in respiration evokes a change in the blood oxygen with delay of approximately $13 - 15$ seconds.

# 6 Conclusion

The extension to time series with several simultaneously measured variables of the nonlinearity test, which combines the redundancy – linear redundancy approach with the surrogate data technique, was proposed. Various types of the redundancies, which can be defined in order to test specific dependence structures between/among (groups of) variables, were demonstrated. Null hypotheses of multivariate linear stochastic processes were tested using the multivariate surrogate data. The linear redundancies were used in order to avoid spurious results due to imperfect surrogates.

Using numerically generated data of know origin, the method was demonstrated to be able to distinguish correctly linear and nonlinear multivariate time series. Examples of several types of the redundancies, evaluated as functions of time lags, were presented to illustrate how the redundancies can be used to study specific dependence structures in multivariate series. In real-data examples, nonlinearity was detected in meteorological and physiological multivariate data. Simple time-lag dependence of the redundancy revealed an interesting phenomenon in the relation between the atmospheric temperature and pressure (a lag between linear and nonlinear underlying mechanisms, which both, however, are driven by the same oscillatory process). In the physiological example, the conditional redundancy was used to evaluate the relation (again as a function of the time lag) between respiration and blood oxygen saturation eliminating the influence of the heart rate.

In summary, the proposed method, which combines the redundancy – linear redundancy approach with the surrogate data technique, is not only a reliable method for detection of nonlinearity in multivariate time series, but can also bring further information about specific relations between/among variables under study and about changes in these relations in dynamics of underlying processes.

**Acknowledgements**

The author would like to thank D. Novotná for providing the meteorological data and K. Eben for his comments on the manuscript.

This study is supported by the Grant Agency of the Czech Republic (grant No. 201/94/1327) and by the Academy of Sciences of the Czech Republic (grant No. 230404).

# References

[1] D.T. Kaplan and L. Glass, *Phys. Rev. Lett.* **68 (4)** (1992) 427-430.

[2] M.B. Kennel and S. Isabelle, *Phys. Rev. A* **46** (1992) 3111-3118.

[3] J. Theiler, S. Eubank, A. Longtin, B. Galdrikian and J.D. Farmer, *Physica D* **58** (1992) 77-94.

[4] M. Paluš, V. Albrecht and I. Dvořák, *Phys. Lett. A* **175** (1993) 203-209.

[5] M. Paluš, In: [18], pp. 387-413.

[6] M. Paluš, *Physica D* **80** (1995) 186-205.

[7] M. Paluš and D. Novotná, *Phys.Lett. A,* **193** (1994) 67-74.

[8] H. Tong, *Nonlinear Time Series Analysis: A Dynamical Systems Approach,* (Oxford University Press, Oxford, 1990).

[9] D.R. Rigney, A.L. Goldberger, W.C. Ocasio, Y. Ichimaru, G.B. Moody and R.G. Mark, In: [18], pp. 105-129.

[10] A. Čenys, G. Lasiene and K. Pyragas, *Physica D* **52** (1991) 332-337.

[11] M.L. Green and R. Savit, *Physica D* **50** (1991) 512-544.

[12] R. Savit and M.L. Green, *Physica D* **50** (1991) 9511-116.

[13] A. Destexhe, J.A. Sepulchre and A. Babloyantz, *Phys. Lett. A* **132** (1988) 101-106.

[14] I. Dvořák, *Phys. Lett. A* **151** (1990) 225.

[15] M. Paluš, I. Dvořák and I. David, *Physica A* **185** (1992) 433-438.

[16] M. Paluš, I. Dvořák and I. David, In: I. Dvořák and A.V. Holden (eds.), *Mathematical Approaches to Brain Functioning Diagnostics,* (Manchester University Press, Manchester & New York, 1991), pp. 369-385.

[17] J. Wackermann, D. Lehmann, I. Dvořák and C.M. Michel, *EEG Clin. Neurophysiol.* **86** (1993) 193-198.

[18] A.S. Weigend and N.A. Gershenfeld, (eds.), *Time Series Prediction: Forecasting the Future and Understanding the Past,* Santa Fe Institute Studies in the Sciences of Complexity, Proc. Vol. XV (Addison–Wesley, Reading, Mass., 1993)

[19] D. Prichard and J. Theiler, *Phys. Rev. Lett.* **73** (1994) 951-954.

[20] S.D. Morgera, *IEEE Trans. on Systems, Man, and Cybernetics* **SMC-15 No. 5** (1985) 608-619.

[21] T.M. Cover and J.A. Thomas, *Elements of Information Theory* (J. Wiley & Sons, New York, 1991).

[22] E.N. Lorenz, *J. Atmos. Sci.* **20** (1963) 130-141.

[23] W.H. Press, B.P. Flannery, S.A. Teukolsky and W.T. Vetterling, *Numerical Recipes: The Art of Scientific Computing* (Cambridge Univ. Press, Cambridge 1986).